\documentclass[prb,twocolumn,10pt,aps,showpacs,groupedaddress]{revtex4-1}
\usepackage{graphicx}
\usepackage{amsmath}
\usepackage{amssymb}
\usepackage{bm}
\usepackage{dcolumn}
\usepackage{soul} 
\usepackage{psfrag}
\usepackage{float}
\usepackage{subeqnarray}
\usepackage{xcolor}

\usepackage{epstopdf}

\usepackage{url}
\usepackage[colorlinks=true,linkcolor=blue,citecolor=blue,urlcolor=blue,breaklinks]{hyperref}

\usepackage{placeins}

\begin{document}
     \nocite{apsrev41control}
\author{L.-F. Zhang}
\author{L. Covaci}
\author{M. V. Milo\v{s}evi\'{c}}\email{milorad.milosevic@uantwerpen.be}
\affiliation{Departement Fysica, Universiteit Antwerpen,
Groenenborgerlaan 171, B-2020 Antwerpen, Belgium}
\title{Topological phase transitions in small mesoscopic chiral $p$-wave superconductors}

\begin{abstract}
Spin-triplet chiral $p$-wave superconductivity is typically described by a two-component order parameter, and as such is prone to unique emergent effects when compared to the standard single-component superconductors. Here we present the equilibrium phase diagram for small mesoscopic chiral $p$-wave superconducting disks in the presence of magnetic field, obtained by solving the microscopic Bogoliubov-de Gennes equations self-consistently. In the ultra-small limit, the cylindrically-symmetric giant-vortex states are the ground state of the system. However, with increasing sample size, the cylindrical symmetry is broken as the two components of the order parameter segregate into domains, and the number of fragmented domain walls between them characterizes the resulting states. Such domain walls are topological defects unique for the $p$-wave order, and constitute a dominant phase in the mesoscopic regime. Moreover, we find two possible types of domain walls, identified by their chirality-dependent interaction with the edge states. 
\end{abstract}

\pacs{74.78.Na, 74.20.Rp, 74.25.Dw}

\maketitle

\section{Introduction}\label{sec:1}

Superconductors described by multi-component order parameters have drawn a lot of attention over the last few decades\cite{tanaka_multicomponent_2015-1, lin_ground_2014,milosper}.  They exhibit many interesting properties that are not possible in the conventional single-component superconductors, such as collective Leggett modes\cite{lin_massless_2012}, fractional vortices\cite{babaev_vortices_2002}, skyrmionic knotted solitons\cite{babaev_hidden_2002}, phase solitons\cite{tanaka_soliton_2001, babaev_hidden_2002, lin_phase_2012}, and hidden criticality \cite{komendova_twoband_2012}, to name a few.

Of particular interest are the spin-triplet chiral $p$-wave superconductors with the order parameter of type $\Delta_{\pm}(p) \sim  p_x \pm i p_y$\cite{kallin_chiral_2012, kallin_chiral_2016}.  Such symmetries can be realized in the A phase of superfluid $^3$He\cite{volovik_universe_2009} and may be attributed to the layered ruthenate superconductor Sr$_2$RuO$_4$\cite{maeno_superconductivity_1994}.  Due to the extra degree of freedom of spin-triplet states, the order parameter is a multi-component one.  In addition, such complex structure of the order parameter breaks the time-reversal symmetry (TRS), indicating that Cooper pairs carry internal angular momentum.  For this reason, chiral $p$-wave superconductors can support rich topological defect states with exotic physical properties.  For example, a vortex exhibits different properties depending on whether its vorticity is parallel or anti-parallel to the internal angular momentum of the Cooper pairs\cite{matsumoto_chiral_1999, sauls_vortices_2009, yokoyama_chirality_2008}.  Further, one finds that domains of different chiralities, namely $p_x + i p_y$ and $p_x - i p_y$, are degenerate in energy and therefore can coexist in the ground state, separated by a domain wall - another type of topological defect unique to $p$-wave superconductivity\cite{matsumoto_quasiparticle_1999, serban_domain_2010, becerra_multichiral_2016-1}.  Such domain wall is attractive for half-quantum vortices\cite{garaud_skyrmionic_2012}, and when enclosed it can form a so-called coreless vortex, with skyrmionic topological properties\cite{fernandez_becerra_vortical_2016, zhang_electronic_2016, garaud_properties_2015}.

Another important aspect of chiral $p$-wave superconductivity is its non-trivial topological order\cite{kallin_chiral_2016}, analogous to that of the Moore-Read state for quantum Hall systems at $5/2$ filling\cite{read_paired_2000}.  A consequence of that topological order is the existence of Majorana zero modes\cite{qi_topological_2011-1}, which obey non-Abelian statistics and hold promise for realization of a topological quantum computer.  A well-known example is that the half-quantum vortex in a chiral $p$-wave superconductor supports a single Majorana zero mode at its core.\cite{ivanov_non-abelian_2001}  However, the half-quantum vortex is thermodynamically unfavorable because its energy diverges logarithmically with the size of the system due to unscreened spin currents. Therefore, one possible way to stabilize such half-quantum vortices is to employ mesoscopic confinement.  The evidences for the existence of half-quantum vortices in a mesoscopic Sr$_2$RuO$_4$ ring\cite{jang_observation_2011}, as well as in trapped superfluid $^3$He\cite{autti_observation_2016}, have recently been reported.

Mesoscopic superconducting systems, whose dimensions are comparable to the penetration depth and the coherence length, often serve as a platform to investigate the fundamental physics of topological defect states. Vortex states in confined conventional superconductors have been well studied over the past few decades\cite{geim_phase_1997, schweigert_vortex_1998, xu_magnetic_2008, berdiyorov_confinement_2009, zhang_unconventional_2012}, with emphasis on their dependence on the size and geometry of the sample. For example, a coalescence of a multi-vortex state into a giant vortex (one vortex but carrying multiple flux quanta) under influence of mesoscopic confinement was predicted and observed experimentally\cite{schweigert_vortex_1998, kanda_experimental_2004, cren_vortex_2011}.  In multi-component superconductivity, mesoscopic samples are expected to stabilize fractional vortices\cite{chibotaru_thermodynamically_2007}, which are thermodynamically unfavorable in bulk samples. However, the emergent states in mesoscopic chiral $p$-wave superconductors are still under debate. For example, Ref.~\onlinecite{huo_field-induced_2012} and Ref. \onlinecite{huang_phase_2012} show contradictory results for small mesoscopic $p$-wave disks, even in the absence of external magnetic field. For that reason, in this paper we study possible states in small mesoscopic chiral $p$-wave superconducting disks by solving the microscopic Bogoliubov-de Gennes (BdG) equations self-consistently, in two dimensions, without any unnecessary assumption. We find that, in contrast to the vortex states always being the ground state in ultimately small disks, domain-wall states become the ground state in larger mesoscopic samples. These domain walls appear upon splitting of the composite vortex, which has coinciding vortex (or anti-vortex) cores in each component of the order parameter. In terms of symmetry breaking, this phase transition is similar to the one between giant vortex states and multi-vortex states in $s$-wave superconductors. These novel states made of domain walls and their interplay with vortices in the mesoscopic limit are unique to $p$-wave order, and therefore relevant to several recently realized systems with $p$-wave-like topological superconductivity\cite{fu_superconducting_2008, mourik_signatures_2012, sau_generic_2010, bernardo_p-wave_2017}. Arguably, our results are most relevant to the nanoscale Pb/Co/Si(111) system of Ref. \onlinecite{menard_two-dimensional_2016}, in which the chiral edge modes were observed and where the dependence of the ground state on the size of the system and applied magnetic field can be directly explored, with an eye on stabilization and observation of the Majorana bound states within the emergent topological defects.

The paper is organized as follows. In Section \ref{sec:2} we introduce our theoretical (Bogoliubov-de Gennes) formalism for $p$-wave superconductors, allowing for non-cylindrically symmetric states. In Section \ref{sec:3} we present the results of our simulations, reporting the equilibrium phase diagram of emergent topological-defect states as a function of the applied field and the size of the mesoscopic $p$-wave system, with thorough investigation of all found states and transitions between them. Our findings are summarized in Section \ref{sec:4}.

\section{Theoretical formalism}\label{sec:2}

The order parameter with chiral $p$-wave pairing symmetry can be expressed as
\begin{equation}\label{Eq:OPvpm}
\mathbf{\Delta}(\mathbf{r},\mathbf{k})=\Delta_+(\mathbf{r})Y_+(\mathbf{k})+\Delta_-(\mathbf{r})Y_-(\mathbf{k}).
\end{equation}
Here $\Delta_{\pm}(\mathbf{r})$ are the spatial $p_x \pm ip_y$-wave order parameters and $Y_{\pm}(\mathbf{k})=(k_x \pm ik_y)/k_F$ are the pairing functions in relative momentum space.  The spinless BdG equations are written as \cite{matsumoto_vortex_2001}:
\begin{equation}\label{Eq:BdG}
  \begin{bmatrix}
    H_e(\mathbf{r}) & \Pi(\mathbf{r}) \\
    -\Pi^*(\mathbf{r}) & -H_e^*(\mathbf{r})
  \end{bmatrix}
  \begin{bmatrix}
    u_n(\mathbf{r}) \\
    v_n(\mathbf{r})
  \end{bmatrix}
  = E_n
  \begin{bmatrix}
    u_n(\mathbf{r}) \\
    v_n(\mathbf{r})
  \end{bmatrix},
\end{equation}
where
\begin{equation}\label{Eq:He}
H_e(\mathbf{r})= \frac{1}{2m_e}[\frac{\hbar}{i}\nabla-\frac{e}{c}\mathbf{A}(\mathbf{r})]^2-E_F
\end{equation}
is the single-particle Hamiltonian with $m_e$ being the electron mass, $E_F$ the Fermi energy and $\mathbf{A}(\mathbf{r})$ the vector potential. We use the gauge $\nabla \cdot \mathbf{A} = 0$.  For simplicity, we consider a cylindrical two dimensional Fermi surface.  In addition, the contribution of the supercurrent to the total magnetic field can be neglected in thin superconducting samples\cite{zhang_electronic_2016}, resulting in a vector potential of form $\mathbf{A}(\mathbf{r}) = \frac{1}{2}H_0 r \mathbf{e}_{\theta}$, with the applied magnetic field $\mathbf{H} = H_0 \mathbf{e}_{z}$. Term $\Pi(\mathbf{r})$ is written as
\begin{equation}\label{Eq:BigPi}
\Pi(\mathbf{r}) = -\frac{i}{k_F} \sum_{\pm} [\Delta_{\pm}\square_{\pm} + \frac{1}{2}(\square_{\pm}\Delta_{\pm})],
\end{equation}
with $\square_{\pm}=e^{\pm i \theta} (\partial_r \pm \frac{i}{r}\partial_\theta )$ in cylindrical coordinates. $u_n(\mathbf{r})$($v_n(\mathbf{r})$) are electron(hole)-like quasiparticle eigen-functions obeying the normalization condition
\begin{equation}\label{Eq:normuv}
  \int |u_n(\mathbf{r})|^2+|v_n(\mathbf{r})|^2 d\mathbf{r}=1,
\end{equation}
and $E_n$ are the corresponding quasiparticle eigen-energies.  The system is considered to be a disk of radius $R$, therefore, the boundary conditions for the wavefunctions are $u_n(r=R)=0$ and $v_n(r=R)=0$.  The order parameters, $\Delta_{\pm}(\mathbf{r})$, satisfy the self-consistent gap equations
\begin{equation}\label{Eq:DxDy}
  \begin{split}
\Delta_{\pm}(\mathbf{r}) = &-i\frac{g}{2k_F}\sum_{E_n<\hbar \omega_D} [v_n^*(\mathbf{r})\square_{\mp}u_n(\mathbf{r})- \\ & u_n(\mathbf{r})\square_{\mp}v_n^*(\mathbf{r})]\times [1-2f(E_n)],
  \end{split}
\end{equation}
where $k_F=\sqrt{2mE_F/\hbar^2}$ is the Fermi wave number, $g$ the superconducting coupling strength and $f(E_n)=[1+\exp(E_n/k_B T)]^{-1}$ is the Fermi-Dirac distribution function.  The summation in Eq.~(\ref{Eq:DxDy}) is over all the quasiparticle states with energies in the Debye window, $\hbar \omega_D$.  Due to the $p_x \pm ip_y$ symmetry, the angular momentum can take the values $\pm1$, and the phase winding numbers $L_{\pm}$ of $\Delta_{\pm}$ always preserve the relative relation $L_-=L_++2$.


To solve the previous equations, we expand the quasiparticle wavefunctions, $u_n(\mathbf{r})$ and $v_n(\mathbf{r})$, in terms of a complete orthonormal basis set:
\begin{equation}\label{Eq:Bessel1}
\begin{pmatrix}
u_n(\mathbf{r})\\
v_n(\mathbf{r})
\end{pmatrix}
=
\sum_{\mu j}
\begin{pmatrix}
c^n_{\mu j} \\
d^n_{\mu j}
\end{pmatrix}
\varphi_{j\mu}(r,\theta),
\end{equation}
where $c^n_{\mu j}$ and $d^n_{\mu j}$ are coefficients, $\mu \in \mathbb{Z}$ are angular quantum numbers corresponding to the angular momentum operator, and the basis functions 
\begin{equation}\label{Eq:Bessel2}
\varphi_{j\mu}(r)=\frac{\sqrt{2}}{RJ_{\mu+1}(\alpha_{j\mu})}J_{\mu}(\alpha_{j\mu}\frac{r}{R}) \frac{e^{i \mu \theta}}{\sqrt{2\pi}},
\end{equation}
with $J_{\mu}$ the $\mu$th Bessel function and $\alpha_{j\mu}$ the $j$th zero of $J_{\mu}$.  The BdG equations can now be reduced to a matrix eigenvalue problem.  Here, we do not impose the cylindrical symmetry on the order parameters.  Therefore, $\Delta_{\pm}$ have the general form $\Delta_{\pm} = \sum_m \Delta_{\pm}(r)e^{im\theta}$.   

To construct the ground-state phase diagram, we calculate the free energy at each self-consistent iteration step (see Appendix~\ref{Ap:sec:FE} for more details) as 
\begin{equation}\label{Eq:G_all}
  \begin{split}
    \mathcal{G} =& \sum_{n} \Big \{ 2E_nf_n -2 E_n \int d\mathbf{r} |v_n|^2 \\
&-\int d\mathbf{r}  \Big [ u_n^* \tilde{\Pi} v_n f_n + v_n \tilde{\Pi} u_n^* (1-f_n) \Big ] \Big \} -TS.
    \end{split}
\end{equation}
Here we define $\tilde{\Pi} = 2\Pi-\Pi'$, where $\Pi$ is calculated using Eq.~\eqref{Eq:BigPi} with the initial input of $\Delta_{\pm}$ but in $\Pi'$ the order parameters $\Delta_{\pm}$ are calculated according to Eq.~\eqref{Eq:DxDy}.  Once the self-consistence loop converges, we obtain $\tilde{\Pi} = \Pi$. The magnetic energy has been neglected since we consider a thin sample.  For convenience, we define the superconducting free energy density $G$ with respect to the corresponding normal state one, i.e.,
\begin{equation}\label{Eq:G_den}
G = (\mathcal{G} - \mathcal{G}_N)/A,
\end{equation}
where $\mathcal{G}_N$ is the free energy of the normal state (i.e. $\Delta_{\pm} \equiv 0$) and $A=\pi R^2$ is the area of the disk sample.  The corresponding bulk superconducting free energy density at zero temperature ($T=0$) in absence of magnetic field ($\phi=0$) is denoted as $G_0$.

Without particular loss of generality, and taking into account numerical convenience, the parameters used in this paper are $E_F=\hbar \omega_D \approx 27 \Delta_0$, resulting in (1) $k_F\xi_0 \approx 18$, where $\xi_0=\hbar v_F/\pi\Delta_0$ is the BCS coherence length at zero temperature, with $v_F$($k_F$) the Fermi velocity (wave number) and $\Delta_0$ the bulk order parameter amplitude at zero temperature; (2) $\Delta_0/k_BT_c = 1.76$ (weak coupling regime), where $k_B$ is the Boltzmann constant. The considered temperature is $T=0.1T_c$, beyond the applicability range of the Ginzburg-Landau-based models.

\section{Results}\label{sec:3}
\begin{figure}
  \centering
  \includegraphics[width=\columnwidth]{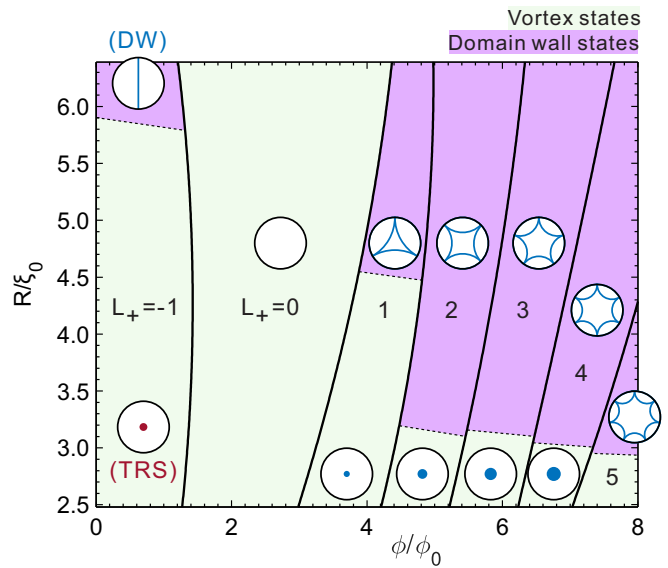}
  \caption{(Color online) The (magnetic flux, size) ground-state phase diagram for a mesoscopic $p$-wave disk. Magnetic flux $\phi$ through the sample is in units of flux quantum $\phi_0=\frac{hc}{2e}$ and radius $R$ of the sample is in units of BCS coherence length $\xi_0$.  In small disks, the ground state are (giant) composite vortex states with cylindrical symmetry, characterized by the sequentially increasing vorticity $L_+$.  With increasing size the system undergoes symmetry-breaking phase transitions from composite vortex states to the chiral domain-wall states (except for the vortex-free state $L_+=0$). The number of chiral domain walls in a given state corresponds to $L_++2$. Dashed lines indicate the second-order phase transitions and solid lines the first-order ones.}
  \label{phasediagram}
\end{figure}

In this study, we focus on small mesoscopic $p$-wave superconducting disks exposed to perpendicular magnetic field, to highlight the unique phase transitions in that regime. The ground state of the system is obtained by comparing the free energy density $G$ of all found states.  The solutions are obtained by using different initial conditions for $\Delta_{\pm}(\mathbf{r})$, such as spatially randomized values (field-cooled conditions), different winding numbers, and/or different vortex configurations. We put forward our main results in Fig.~\ref{phasediagram}, summarizing the equilibrium phase diagram as a function of the radius of the sample $R$, and the magnetic flux $\phi$ threading the sample.  

Contrary to what many would expect, the ground state is not vortex-free in absence of magnetic field. As seen in Fig.~\ref{phasediagram}, the ground state at zero field is a vortex state with winding numbers $L_\pm = \mp1$ in the two components of the order parameter. The vortex-free state, corresponding to the conventional Meissner state of $s$-wave superconductors, stabilizes as the ground state only at higher magnetic field.

\begin{figure}
  \centering
  \includegraphics[width=\columnwidth]{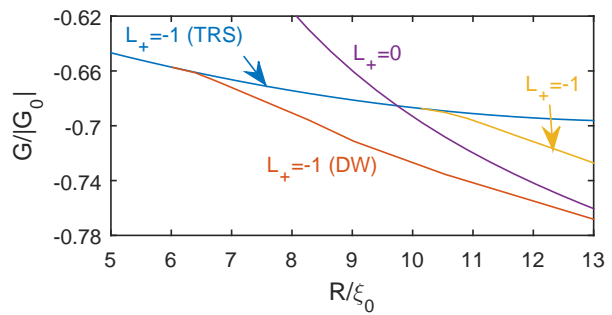}
  \caption{(Color online) The superconducting free energy density $G$ of the superconducting states as a function of the sample size ($R$), in absence of magnetic field ($\phi=0$). The $L_+=-1$ domain wall (DW) state replaces the time-reversal-symmetric (TRS) $L_+=-1$ vortex state as the ground state for $R\gtrsim6\xi_0$.  Note that the vortex-free state with $L_+=0$ becomes the ground state in much larger disks, while the anti-parallel vortex state with $L_+=-1$ (in which $\Delta_-$ component is nearly completely suppressed) only exists as a metastable state.}
  \label{freeenergy}
\end{figure}

We have actually checked that the ground state in the bulk sample with same parameters is the vortex-free state with $L_+ = 0$ ($L_-=L_++2$ holds in all cases, and will be omitted from here on). Why is then the vortex state with $L_+ = -1$ the ground state at zero field for mesoscopic samples? To explain this, we recall that chiral $p$-wave superconductors host edge states with an edge current\cite{suzuki_spontaneous_2016, bouhon_current_2014} due to their topological nature. When decreasing the size of the sample ($R$), the edge states overlap and their interaction increases, causing destruction of superconductivity in the vicinity of the sample center, very similar to the normal vortex core.  Due to this, as seen from Fig.~\ref{freeenergy}, the Gibbs free energy density of the vortex-free state with $L_+ = 0$ increases strongly with $R$ decreasing. In contrast, the vortex state with $L_+ = -1$ in ultra small disks and at zero field exhibits lower energy, and preserves time-reversal-symmetry (TRS). In this case, both components of the order parameter not only satisfy cylindrical symmetry, i.e. $\Delta_{\pm} = |\Delta_{\pm}(r)|e^{iL_{\pm}\theta}$, but also have the same spatial distribution, i.e. $|\Delta_{+}(r)| = |\Delta_{-}(r)|$.  In other words, there is an anti-vortex at the center of the sample in one component, and a vortex at the very same place in the other component. Their supercurrents ideally cancel each other, making this phase stable at zero field.  As a result, the $L_+ = -1$ (TRS) state is fostered by the interaction between the edge states, and the free energy of this state does not increase as fast as the energy of the vortex-free state with $R$ decreasing.  As seen from Fig.~\ref{freeenergy}, the $L_+ = -1$ (TRS) state is the ground state in a wide range of small disks ($R\lesssim6\xi_0$). 

Note that the so-called anti-parallel vortex state (with $L_+ = -1$, but nearly completely suppressed $\Delta_-$ component of the order parameter) does not exist in such a small disk. Although it stabilizes in larger samples, it remains metastable (i.e. with higher energy, as shown in Fig. \ref{freeenergy}), even for non-zero external magnetic field.

\begin{figure}
  \centering
  \includegraphics[width=\columnwidth]{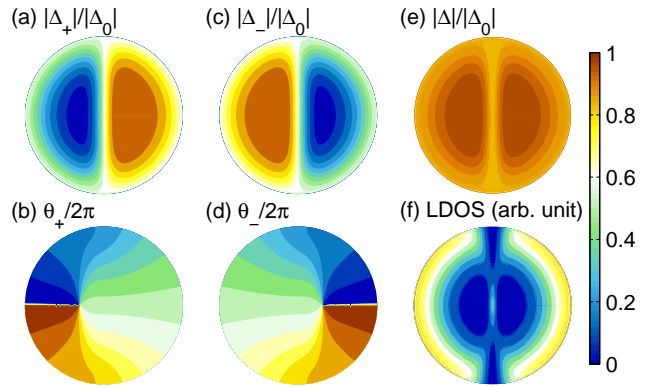}
  \caption{(Color online) Characterization of the $L_+=-1$ domain-wall state.  For the sample of size $R=13\xi_0$ and zero magnetic field, we show the spatial distribution of the two components of the order parameter $|\Delta_{\pm}|$ (a,c) and their phase $\theta_{\pm}$ (b,d), respectively. Panel (e) shows the spatial distribution of the total order parameter $|\Delta|= \sqrt{|\Delta_+|^2+|\Delta_-|^2}$ and (f) is the corresponding zero-bias LDOS.}
  \label{Vtx-1}
\end{figure}

When the size of the sample $R$ is increased, the confinement weakens. In this case, we find that in the $L_+ = -1$ (TRS) state the cores of the anti-vortex in $\Delta_{+}$ and the vortex in $\Delta_{-}$ separate from each other, leading to broken cylindrical symmetry.  As an illustrative example, we show the spatial profile of the order parameter after the formation of such state in Fig.~\ref{Vtx-1}(a,c,e). As seen there, $\Delta_{+}$ and $\Delta_{-}$ segregate, and $|\Delta_{+}|$ becomes mirror symmetric with $|\Delta_{-}|$, as the anti-vortex shifts to the left side of the sample while the vortex shifts to the right side [see the plots of the phase of the respective components of the order parameter in Fig.~\ref{Vtx-1}(b,d)]. This results in a clear chiral domain wall between $\Delta_{+}$ and $\Delta_{-}$, passing through the center of the sample. The total order parameter, $|\Delta|$, is suppressed at the domain wall [shown in Fig.~\ref{Vtx-1}(e)]. In addition, as a topological defect, the domain wall carries low-lying bound states passing Fermi energy, leading to low-lying LDOS distributions at the domain wall\cite{zhang_electronic_2016}. Fig.~\ref{Vtx-1}(f) shows the zero-bias LDOS for the $L_+ = -1$ (DW) state in the sample with $R=13\xi_0$. The domain wall yields a maximum in LDOS near the sample center.  Near the sample edge, the domain wall causes suppression of the LDOS due to the interference effect between the domain-wall bound states and the edge bound states. The latter ones yield enhanced LDOS everywhere else near the sample edges. With increasing external magnetic field, the domain wall shifts to either left or right depending on the polarity of the applied field.

The $L_+ = -1$ (DW) state replaces the $L_+ = -1$ (TRS) state as the ground state in larger samples, with radius beyond $\approx6\xi_0$ (as seen in Fig.~\ref{freeenergy}, $L_+ = -1$ (DW) state attains lower free energy with respect to $L_+ = -1$ (TRS) state for $R>6\xi_0$). The phase transition between $L_+ = -1$ (DW) and $L_+ = -1$ (TRS) state is of second order and therefore fully reversible, either as a function of size or as a function of applied magnetic field (corresponding to crossing the dashed line in Fig. \ref{phasediagram}).

Thus, to summarize the discussion so far, there are three stable states at zero field for a mesoscopic chiral $p$-wave superconductor, namely i) the $L_+ = -1$ (TRS) vortex state; ii) the one domain wall state with $L_+ = -1$; and iii) the vortex-free state with $L_+ = 0$. These states respectively replace one another in the ground state of the system, as sample is made progressively larger.

With increasing magnetic field, we find similar relationship between the vortex states and the domain-wall states for given vorticity. In the limit of very small samples, the composite (giant) vortex of successively increasing vorticity is the cylindrically-symmetric ground state of the system, in agreement with previous Ginzburg-Landau investigations\cite{huang_phase_2012}. Note however, that this behavior is not captured in Ref.~\onlinecite{huo_field-induced_2012}, using a BdG formalism, where an erroneous phase diagram is presented. This being aside the main message of our present work, we provide more details on the comparison of our results to those of Ref. \onlinecite{huo_field-induced_2012} in Appendix~\ref{Ap:sec:Disk}.
\begin{figure}
  \centering
  \includegraphics[width=\columnwidth]{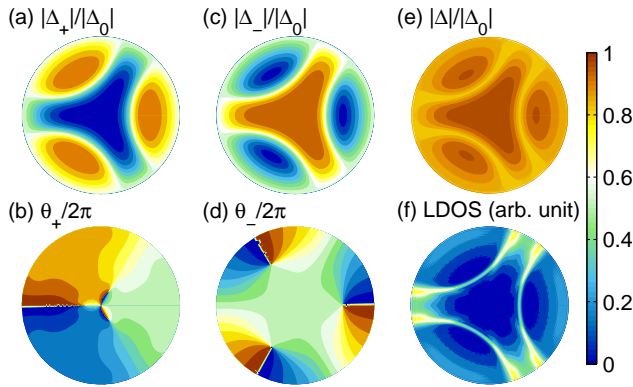}
  \caption{(Color online) Characterization of the $L_+=1$ (three) domain-wall state, for the sample with $R=13\xi_0$ and $\phi=4.8\phi_0$. Displayed quantities are the same as in Fig.~\ref{Vtx-1}.}
  \label{Vtx1}
\end{figure}

The main focus of our present investigation is the topological phase transition occurring with increasing size of mesoscopic chiral $p$-wave superconductors, during which the two components of the order parameter segregate, and the composite vortex states are replaced by domain wall states as the ground state of the system, for any applied magnetic field. For example, the vortex state with $L_+ = 1$ obeys cylindrical symmetry in ultra small samples ($R\lesssim4.5\xi_0$), but develops into the three-domain-wall state in larger samples, shown in Fig.~\ref{Vtx1}. More precisely, during this transition the giant vortex in $\Delta_-$ with winding number $L_-=3$ splits into a multi-vortex state.  Meanwhile, the vortex state in $\Delta_+$ with $L_+ =1$ hosts a vortex-antivortex molecule with net vorticity 1 (in order to best complement the symmetry of $\Delta_-$, instead of a simple vortex, a giant anti-vortex with $L=-2$ is formed at the disk center surrounded by three vortices). At the domain walls, the total order parameter is depleted and LDOS exhibits zero bias peaks. 

\begin{figure}
  \centering
  \includegraphics[width=\columnwidth]{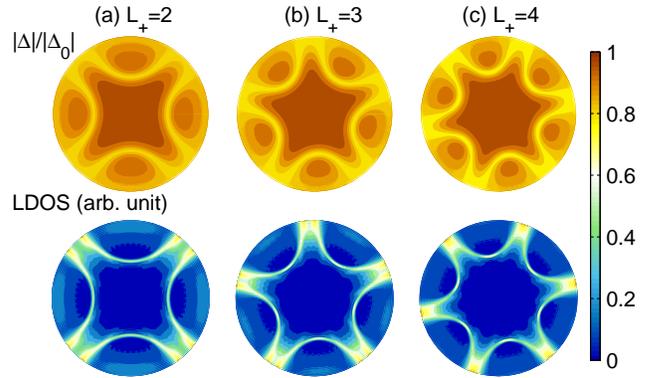}
  \caption{(Color online) The order parameter distribution $|\Delta|$ (upper panels) and the corresponding zero-bias LDOS (lower panels) of the domain-wall states with $L_+=$2, 3, and 4 (respectively from left to right).}
  \label{Vtx2t4}
\end{figure}

To generalize our findings to all fields, the $L_+$ vorticity sequentially increases with applied magnetic field, and the number of domain walls for given $L_+$ vorticity matches $L_-=L_++2$. $\Delta_+$ consists of spread $L_++2$ vortices and a centered giant anti-vortex with $L=-2$, while $\Delta_-$ consists of $L_++2$ vortices. Spatial distributions of the order parameter (directly verifiable by e.g. STM) for states of higher vorticity are shown in Fig.~\ref{Vtx2t4}, together with the corresponding zero-bias LDOS, all exhibiting low-lying states inside the domain walls. In the present consideration we neglected the self-field of the superconductor, but it is trivial to conclude that the self-field of domain-wall states would be focused at the domain walls, offering a route for direct verification in magnetic-probe microscopy\cite{khotkevych_scanning_2008, curran_search_2014}. The range of sample sizes and the applied field needed for stability of each of the domain-wall states in the ground state is shown in Fig.~\ref{phasediagram}. We note that the domain-wall states dominate the composite vortex states as magnetic field increases, i.e. the cylindrically-symmetric states are not sufficient to describe the ground state at larger magnetic field even in very small $p$-wave samples. 

As a last remark, we notice that the behavior of the zero-bias LDOS for the $L_+ = -1$ domain-wall state is different from the other domain-wall states at higher field [cf. Fig. \ref{Vtx-1}(e) and lower panels of Fig.~\ref{Vtx2t4}]. In the $L_+ = -1$ state, the domain wall suppresses the edge states in the sample, while the domain walls in higher-vorticity states enhance the LDOS peaks at the sample edge. This is due to the fact that in $L_+ = -1$ state the domain wall separates a vortex in $\Delta_{-}$ and an antivortex in $\Delta_{+}$, while in other states the domain wall separates vortices in both $\Delta_{-}$ and $\Delta_{+}$.  As a result, we identify two types of domain walls, exhibiting chirality-dependent effects on the edge states\footnote{For a similar polarity-dependent effect on edge states due to nearby (anti)parallel vortex, see Ref. \onlinecite{yokoyama_chirality_2008}.}.

\section{Conclusions}\label{sec:4}

In conclusion, motivated by recent experimental efforts, we reported the ground-state phase diagram in small-to-intermediate mesoscopic chiral $p$-wave superconducting samples exposed to perpendicular magnetic field. For this study, we employed the self-consistent Bogoliubov-de Gennes numerical formalism, where we went beyond the approximations of similar earlier works. In ultra small samples, due to strong confinement, the ground states are the (giant) composite vortex states obeying cylindrical symmetry, in agreement with earlier results in literature\cite{huang_phase_2012} and conventional wisdom for small superconductors. However, in samples larger than few coherence lengths, the domain-wall states replace the vortex states as the ground state of the system. These domain walls mark the segregation of the two components of the order parameter and the vortices therein, and their number increases with increasing magnetic field. We also reveal two types of domain walls in a chiral $p$-wave superconductor, distinguished by their chirality-dependent effect on the edge states in LDOS. These domain walls can be directly visualized by scanning tunneling microscopy as locally suppressed order parameter and correspondingly boosted bound states in LDOS, or by magnetic-probe microscopy as locally focused self-field of the sample. These predictions are directly relevant to recent realizations of two-dimensional topological superconductivity, as in for example atomically thin Pb/Co on Si(111), where the size of the $p$-wave system is determined by the size of the underlying Co cluster\cite{menard_two-dimensional_2016}. On theoretical grounds, our results translate to other superconducting systems with a multicomponent order parameter, where similar topological transitions from overlapping to segregated components can be expected and emergent physics can be significantly richer than in standard single-component superconductors.


\appendix


\section{\label{Ap:sec:FE} The free energy}

In this Appendix, we derive the expression of the Gibbs free energy for the chiral $p$-wave superconducting state. Throughout the derivation, the Cartesian system of coordinates is used.  Subsequently, we apply the obtained expression for the free energy to the homogeneous system as a quick check of the validity of our approach.  

\subsection{\label{Ap:subsec:FE} Derivation of the free-energy equation}

The Bogoliubov-de Gennes (BdG) equations for a chiral spin-triplet p-wave superconductor\cite{furusaki_spontaneous_2001}, with the order parameter described by the vector $\vec{d}(\vec{k})=\hat{z}\frac{\Delta}{k_F}(k_x \pm i k_y)$, are written as:
\begin{equation}\label{Ap:Eq:BdG}
  \begin{bmatrix}
    H_e & \Pi(\mathbf{r}) \\
    -\Pi^*(\mathbf{r}) & -H_e^*
  \end{bmatrix}
  \begin{bmatrix}
    u_n(\mathbf{r}) \\
    v_n(\mathbf{r})
  \end{bmatrix}
  = E_n
  \begin{bmatrix}
    u_n(\mathbf{r}) \\
    v_n(\mathbf{r})
  \end{bmatrix},
\end{equation}
where
\begin{equation}\label{Ap:Eq:He}
H_e=\frac{1}{2m}(\frac{\hbar}{i}\nabla-\frac{e}{c}\mathbf{A})^2-E_F
\end{equation}
is the single-particle Hamiltonian, with $E_F$ the Fermi energy and $\mathbf{A}$ the vector potential. Term $\Pi(\mathbf{r})$ is written as
\begin{equation}\label{Ap:Eq:BigPi}
  \begin{split}
    \Pi(\mathbf{r}) = &-\frac{i}{k_F} \Big \{ \Delta_x(\mathbf{r})\frac{\partial }{\partial x} +i\Delta_y(\mathbf{r})\frac{\partial }{\partial y} \\
    &+ \frac{1}{2} \big [ \frac{\partial \Delta_x(\mathbf{r})}{\partial x} +i\frac{\partial \Delta_y(\mathbf{r})}{\partial y} \big ] \Big \},
  \end{split}
\end{equation}
where $\Delta_x(\mathbf{r})$ and $\Delta_y(\mathbf{r})$ are $p_x$ and $p_y$ components of the order parameter, respectively, $u_n$($v_n$) are electron(hole)-like quasi-particle eigen-wavefunctions, and $E_n$ are the corresponding quasi-particle eigen-energies. The transformation from the $\Delta_{x}$, $\Delta_{y}$ to $\Delta_{\pm}$ is that $\Delta_x = \Delta_+ + \Delta_-$, $\Delta_y = \Delta_+ - \Delta_-$.  In addition, the self-consistent gap equations are
\begin{equation}\label{Ap:Eq:DxDy}
  \begin{split}
    \Delta_x(\mathbf{r}) = &-i\frac{g}{k_F}\sum_{E_n<\hbar \omega_D} [v_n^*(\mathbf{r})\frac{\partial u_n(\mathbf{r})}{\partial x}-u_n(\mathbf{r})\frac{\partial v_n^*(\mathbf{r})}{\partial x}] \\
    & \times [1-2f(E_n)], \\
    \Delta_y(\mathbf{r}) = &-\frac{g}{k_F}\sum_{E_n<\hbar \omega_D} [v_n^*(\mathbf{r})\frac{\partial u_n(\mathbf{r})}{\partial y}-u_n(\mathbf{r})\frac{\partial v_n^*(\mathbf{r})}{\partial y}]\\
    & \times [1-2f(E_n)],
  \end{split}
\end{equation}
where $f(E_n)$ is the Fermi distribution function.  The summations in Eqs.~\eqref{Ap:Eq:DxDy} are over all the quasi-particle states with energies in the Debye window $\hbar \omega_D$.

Next, we derive the Gibbs free energy for a $p$-wave superconductor, first written as
\begin{equation}\label{Ap:Eq:Gibbs}
    \mathcal{G} = \langle H_{eff} \rangle -TS +F_H,
\end{equation}
where $F_H=(\mathbf{B}-\mathbf{H})^2/8\pi$ is the magnetic energy, $TS$ is the energy induced by entropy, and $\langle H_{eff} \rangle$ is the expectation value of the effective Hamiltonian.  According to Ref.~\onlinecite{ketterson_superconductivity_1999}, the effective Hamiltonian for an unconventional superconductor is written as
\begin{equation}\label{Ap:Eq:Heff}
\begin{split}
\langle H_{eff} \rangle = &\underbrace{\int d\mathbf{r} \int d\mathbf{r}' \sum_{\sigma} \big[   \Psi^{\dagger}(\mathbf{r}\sigma) \hat{H}_{e}\delta(\mathbf{r}-\mathbf{r}') \Psi(\mathbf{r}'\sigma) \big]}_{\langle H_e \rangle} \\
&+ \underbrace{\int d\mathbf{r} \int d\mathbf{r}' \Delta(\mathbf{r}, \mathbf{r}') \Psi^{\dagger}(\mathbf{r}\uparrow) \Psi^{\dagger}(\mathbf{r}'\downarrow)}_{\langle \Delta \rangle}\\
&+ \underbrace{\int d\mathbf{r} \int d\mathbf{r}' \Delta^*(\mathbf{r},\mathbf{r}') \Psi(\mathbf{r}'\downarrow) \Psi(\mathbf{r}\uparrow)}_{\langle \Delta^* \rangle} \\
&+ \underbrace{\int d\mathbf{r} \int d\mathbf{r}' W(\mathbf{r},\mathbf{r}')}_{\langle W \rangle},
\end{split}
\end{equation}
where
\begin{equation}\label{Ap:Eq:W}
W (\mathbf{r},\mathbf{r}')= \hat{V} (\mathbf{r},\mathbf{r}') \big \langle \Psi^{\dagger}(\mathbf{r},\uparrow) \Psi^{\dagger}(\mathbf{r}',\downarrow) \big \rangle  \big \langle \Psi(\mathbf{r}',\downarrow) \Psi(\mathbf{r},\uparrow) \big \rangle 
\end{equation}
is a constant term in the mean-field theory needed to match the expectation value of the mean-field Hamiltonian with the one of the many-body Hamiltonian.  For $p$-wave paring\cite{matsumoto_vortex_2001}, the spatial dependence of the order parameter can be written as:
\begin{equation}\label{Ap:Eq:DeltaRX}
\Delta(\mathbf{r},\mathbf{r}')= \frac{i}{k_F}[\Delta_x(\mathbf{R})\partial_{x'} + i\Delta_y(\mathbf{R})\partial_{y'} ]\delta(\mathbf{r}-\mathbf{r}'),
\end{equation}
with $\mathbf{R}=(\mathbf{r}+\mathbf{r}')/2$.

Next, we calculate the first term in Eq.~(\ref{Ap:Eq:Heff}), $\langle H_e \rangle$. First, we introduce the Bogoliubov transformations:
\begin{equation}\label{Ap:Eq:UnitaryTran}
\begin{split}
\Psi^{\dagger}(\mathbf{r},\uparrow) &= \sum_{n} \Big( u^{*}_{n}(\mathbf{r}) \gamma^{\dagger}_{n\uparrow}+v_{n}(\mathbf{r}) \gamma_{n\downarrow} \Big),\\
\Psi(\mathbf{r},\downarrow) &= \sum_{n} \Big( v^{*}_{n}(\mathbf{r}) \gamma^{\dagger}_{n\uparrow}+u_{n}(\mathbf{r}) \gamma_{n\downarrow} \Big).
    \end{split}
\end{equation}
Note that these are slightly different from those used for $s$-wave paring.  The thermal averages of the fermionic $\gamma$ operators are as usual
\begin{equation}\label{Ap:Eq:meangamma}
  \begin{split}
    \langle \gamma_{n\sigma}^\dagger \gamma_{n'\sigma'} \rangle &= \delta_{\sigma\sigma'} \delta_{nn'} f(E_n),\\
    \langle \gamma_{n\sigma} \gamma_{n'\sigma'}^\dagger \rangle &= \delta_{\sigma\sigma'} \delta_{nn'} [1-f(E_n)],\\
    \langle \gamma_{n\sigma} \gamma_{n'\sigma'}  \rangle &=  \langle \gamma_{n\sigma}^\dagger \gamma_{n'\sigma'}^\dagger  \rangle = 0.
  \end{split}
\end{equation}
Substituting Eqs.~(\ref{Ap:Eq:UnitaryTran}) and (\ref{Ap:Eq:meangamma}) into $\langle H_e \rangle$, we obtain
\begin{equation}\label{Ap:Eq:meanHe}
\langle H_e \rangle =2\int d\mathbf{r} \sum_{n} [u_n^*\hat{H_e} u_n f_n + v_n\hat{H_e} v_n^* (1-f_n) ].
\end{equation}
The factor of $2$ is due to the two spin orientations in the electron and the hole sectors.  Substituting Eq.~(\ref{Ap:Eq:BdG}) into Eq.~(\ref{Ap:Eq:meanHe}), yields
\begin{equation}\label{Ap:Eq:meanHe2}
\begin{split}
\langle H_e \rangle = &2\int d\mathbf{r} \sum_{n} \Big \{ E_n|u_n|^2 f_n-E_n|v_n|^2(1-f_n) \\
&- u_n^*\Pi(\mathbf{r}) v_n f_n - v_n\Pi(\mathbf{r}) u_n^* (1-f_n) \Big \}.
    \end{split}
\end{equation}
Here $\Pi$ is calculated using Eq.~(\ref{Ap:Eq:BigPi}) with the initial input of $\Delta_x$ and $\Delta_y$.

Next, we calculate the second term $\langle \Delta \rangle$ in Eq.~(\ref{Ap:Eq:Heff}). Substituting Eq.~(\ref{Ap:Eq:UnitaryTran}) and (\ref{Ap:Eq:meangamma}) into it, we obtain
\begin{equation}\label{Ap:Eq:meanD1}
  \begin{split}
     \langle \Delta \rangle =& \int d\mathbf{r} \int d\mathbf{r}' \Delta(\mathbf{r}, \mathbf{r}') \big [ u^*_n(\mathbf{r})v_n(\mathbf{r}')f_n \\
     &+ v_n(\mathbf{r})u^*_n(\mathbf{r}')(1-f_n)  \big ].
  \end{split}
\end{equation}
Substituting Eq.~(\ref{Ap:Eq:DeltaRX}) into Eq.~(\ref{Ap:Eq:meanD1}) and integrating by parts, yields
\begin{equation}\label{Ap:Eq:meanDfi}
  \begin{split}
    \langle \Delta \rangle = &\int d\mathbf{r} \sum_{n} \Big \{ u_n^*\Pi'(\mathbf{r}) v_n f_n \\
     &+ v_n\Pi'(\mathbf{r}) u_n^* (1-f_n) \Big \}.
  \end{split}
\end{equation}
Here $\Pi'$ indicates that $\Delta_x$ and $\Delta_y$ are calculated from Eq.~(\ref{Ap:Eq:DxDy}) using the quasi-particle states. 
Similarly, the third term in Eq.~(\ref{Ap:Eq:Heff}) becomes $\langle \Delta^* \rangle = \langle \Delta \rangle$ and the last term $\langle W\rangle = - \langle \Delta \rangle$.  In the calculation of the $\langle W\rangle$ term we used the definition of the order parameter, $\Delta(\mathbf{r},\mathbf{r'})=\hat{V} (\mathbf{r},\mathbf{r}') \big \langle \Psi(\mathbf{r}',\uparrow) \Psi(\mathbf{r}',\downarrow) \big \rangle$.

Finally, the Gibbs free energy is written as
\begin{equation}\label{Ap:Eq:G_all}
  \begin{split}
    \mathcal{G} =& \langle H_{eff} \rangle -TS +F_H \\
    =& \sum_{n} \Bigg \{ 2E_nf_n -2 E_n \int d\mathbf{r} |v_n|^2 \\
&-2\int d\mathbf{r}  \Big [ u_n^*\Pi(\mathbf{r}) v_n f_n + v_n\Pi(\mathbf{r}) u_n^* (1-f_n) \Big ] \\
&+ \int d\mathbf{r} \Big [ u_n^*\Pi'(\mathbf{r}) v_n f_n + v_n\Pi'(\mathbf{r}) u_n^* (1-f_n) \Big ] \Bigg \}\\
&-TS+F_H.
    \end{split}
\end{equation}

When the converged $\Delta_x$ and $\Delta_y$ are reached, then $\Pi = \Pi'$ and the Gibbs free energy shown in Eq.~(\ref{Ap:Eq:G_all}) will be further reduced to
\begin{equation}\label{Ap:Eq:Heff_lastep}
\begin{split}
\mathcal{G} =& \sum_{n} \Bigg \{ 2E_nf_n -2 E_n \int d\mathbf{r} |v_n|^2 \\
&-\int d\mathbf{r}  \Big [ u_n^*\Pi(\mathbf{r}) v_n f_n + v_n\Pi(\mathbf{r}) u_n^* (1-f_n) \Big ] \Bigg \}\\
&-TS+F_H.
    \end{split}
\end{equation}
This is the expression we used in the present study. The summation is done over the entire spectrum. Note that there are \textit{no assumptions made} on the energy range of the pairing potential $V(r,r')$, therefore this expression is valid irrespectively of the relation between $\Delta$ and $\hbar \omega_D$. In the weak-coupling BCS limit, i.e. for $\Delta << \hbar \omega_D << E_F$, and for $s$-wave pairing, this expression reduces to the well-known expression:
\begin{equation}
\mathcal{G} = \sum_{n} \left( 2E_nf_n -2 E_n \int d\mathbf{r} |v_n|^2 \right) + \int d\mathbf{r}\frac{|\Delta|^2}{g} -TS + F_H,
\end{equation}
which is another proof of the consistency of our derivation.

\subsection{\label{Ap:subsec:Homo} Free energy of the homogeneous system}

In this subsection, we present the form of the equations in the homogeneous case in order to show that our free energy is correct.  If the expression of the free energy is valid, the calculated energy should always decrease during the self-consistent procedure since the order parameter approaches its true (lowest) equilibrium value.

For the homogeneous case (two-dimensional), we consider a square unit cell of length $W$.  The components of the order parameter are constant, i.e. $\Delta_x(\mathbf{r})= \Delta_x$ and $\Delta_y(\mathbf{r})= \Delta_y$. $u_n(\mathbf{r})$ and $v_n(\mathbf{r})$ can be expanded in terms of plane waves, i.e. as
\begin{equation}
\binom{u_n(\mathbf{r})}{v_n(\mathbf{r})}= \sum_{\mathbf{k}}\frac{e^{i\mathbf{k} \cdot \mathbf{r}}}{W} \binom{u_\mathbf{k}}{v_\mathbf{k}},
\end{equation}
where $\mathbf{k}=(k_x, k_y) = 2\pi(j_x,j_y)/W$ is the momentum, given that $j_x,j_y \in \mathbb{Z}$. In the absence of the magnetic field, $\mathbf{k}$ is a good quantum number and the BdG equations (\ref{Ap:Eq:BdG}) for a given $\mathbf{k}$ are written as:
\begin{equation}\label{Ap:Eq:BdGhomo}
  \begin{bmatrix}
    \epsilon_\mathbf{k} & \Pi_\mathbf{k} \\
    \Pi^*_\mathbf{k} & -\epsilon_\mathbf{k}
  \end{bmatrix}
  \begin{bmatrix}
    u_\mathbf{k} \\
    v_\mathbf{k}
  \end{bmatrix}
  = E_\mathbf{k}
  \begin{bmatrix}
    u_\mathbf{k} \\
    v_\mathbf{k}
  \end{bmatrix},
\end{equation}
where $\epsilon_\mathbf{k} = \hbar^2 \mathbf{k}^2/2m -E_F$ and $\Pi_\mathbf{k} = (k_x\Delta_x+ik_y\Delta_y) /k_F$.  We take the basis functions whose $\epsilon_\mathbf{k}$ satisfy $|\epsilon_\mathbf{k}| < \hbar \omega_D$.  Then, the self-consistent gap equations (\ref{Ap:Eq:DxDy}) are
\begin{equation}\label{Ap:Eq:DxDyhomo}
  \begin{split}
    \Delta_x &= \frac{2g}{k_F} \sum_{E_\mathbf{k} <\hbar \omega_D} k_x u_\mathbf{k} v^*_\mathbf{k} [1-2f(E_\mathbf{k})], \\
    \Delta_y &= -i\frac{2g}{k_F} \sum_{E_\mathbf{k} <\hbar \omega_D} k_y u_\mathbf{k} v^*_\mathbf{k} [1-2f(E_\mathbf{k})].
  \end{split}
\end{equation}
From Eq.~(\ref{Ap:Eq:G_all}), the Gibbs free energy is
\begin{equation}\label{Ap:Eq:G_allhomo}
  \begin{split}
    \mathcal{G} = &\sum_{\mathbf{k}} \Bigg \{ 2E_\mathbf{k}f(E_\mathbf{k}) -2 E_\mathbf{k} |v_\mathbf{k}|^2 \\
&- \Big [ u_\mathbf{k}^* (2\Pi_\mathbf{k}-\Pi'_\mathbf{k}) v_\mathbf{k} f(E_\mathbf{k}) \\
&+ v_\mathbf{k} (2\Pi_\mathbf{k}-\Pi'_\mathbf{k}) u_\mathbf{k}^* (1-f(E_\mathbf{k})) \Big ] \Bigg \} -TS.
    \end{split}
\end{equation}
The superconducting free energy density $G$ is given according to Eq.~(\ref{Eq:G_den}).

\begin{figure}
  \centering
  \includegraphics[width=\columnwidth]{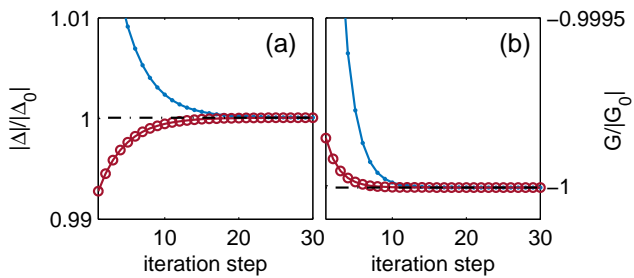}
  \caption{(Color online) (a) The order parameter $\Delta_x$ (which equals $\Delta_y$ in this simulation), and (b) the superconducting free energy density $G$, as a function of the iteration step.  The solid line corresponds to the case of initial order parameter larger than the converged value, while the line with open dots corresponds to the initial order parameter smaller than the converged value.  For both cases the free energy decreases with the iterations, and the final values are the same.}
  \label{Ap:fig:homo}
\end{figure}
Next, we present results for parameters as those used in Ref. \onlinecite{huo_field-induced_2012}. $W=1~\mathrm{{\mu}m}$ is sufficiently large so that results will not change for further enlarged $W$.  Figs.~\ref{Ap:fig:homo} (a) and (b) show the calculated order parameter and the corresponding free energy, respectively,  as a function of iteration steps at $T=0~\mathrm{K}$.  During the iterations, we find that two components of the order parameter are always the same, i.e. $\Delta_x= \Delta_y = \Delta$.  $\Delta$ converges to $0.22~\mathrm{meV}$, which results in the zero temperature coherence length $\xi_0\approx 72~\mathrm{nm}$ ($\xi_0=\hbar v_F /\pi \Delta_0$, with Fermi velocity $v_F=\sqrt{2E_F/m_e}$, Fermi energy $E_F=16.32~\mathrm{meV}$ and zero-temperature order parameter $\Delta_0=0.22~\mathrm{meV}$) and the critical temperature $T_{c,bulk} \approx 1.5 ~\mathrm{K}$.  The free energy shown in Fig.~\ref{Ap:fig:homo}(b) always decreases with the iterations and yields the same result, irrespectively of the initial amplitude of the order parameter being larger or smaller than the converged value. These results prove the reliability of our free energy calculation.

Note that the Gibbs free energy from Eq.~(\ref{Ap:Eq:G_allhomo}) for converged order parameters can be written as:
\begin{equation}\label{Ap:Eq:G_allhomo1}
  \begin{split}
    \mathcal{G} =& \sum_{\mathbf{k}} \Bigg \{ 2E_\mathbf{k}f(E_\mathbf{k}) -2 E_\mathbf{k} |v_\mathbf{k}|^2 \\
&+  \Pi_\mathbf{k} u_\mathbf{k}^* v_\mathbf{k} [1-2f(E_\mathbf{k})] \Bigg \} - TS.
    \end{split}
\end{equation}
We find that
\begin{equation}
\sum_{\mathbf{k}} \Pi_\mathbf{k} u_\mathbf{k}^* v_\mathbf{k} [1-2f(E_\mathbf{k})] = \frac{|\Delta_x|^2+|\Delta_y|^2}{2g} W^2,
\end{equation}
given that all the $\mathbf{k}$-states have the same coupling constant $g$.  Factor $W^2$ is due to the area of the unit cell. Finally, in the homogeneous case the Gibbs free energy can be written as:
\begin{equation}
\mathcal{G} =\sum_{\mathbf{k}} 2E_\mathbf{k}f(E_\mathbf{k}) -2 E_\mathbf{k} |v_\mathbf{k}|^2 + \frac{|\Delta_x|^2+|\Delta_y|^2}{2g} W^2 -TS.
\end{equation}

\section{\label{Ap:sec:Disk} Comparison with other available results in the literature}
\begin{figure}
\centering
\includegraphics[width=\columnwidth]{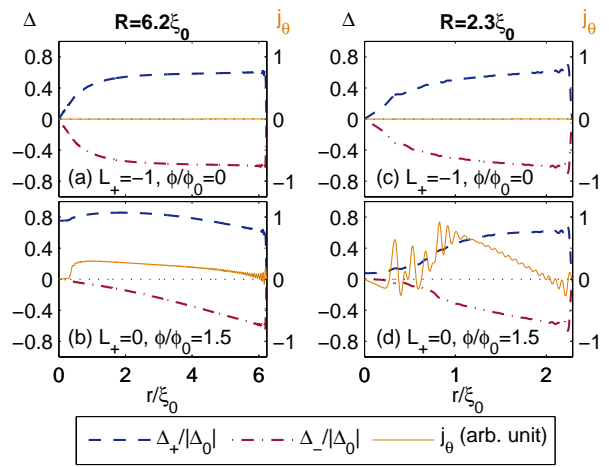}
\caption{(Color online) Spatial distribution of the order parameter components $\Delta_\pm(r)$ and the supercurrent density $j_\theta(r)$, for enforced cylindrical symmetry, in the sample with radius $R= 6.2\xi_0$ for $L_+=-1$ (a) and $L_+=0$ (b), and for the sample with radius $R= 2.3\xi_0$ for $L_+=-1$ (c) and $L_+=0$ (d).} \label{Ap:Fig.1}
\end{figure}

In this Appendix, we discuss in more detail the bottom part of the phase diagram presented in Fig. \ref{phasediagram}, where only cylindrically symmetric states are found. This is needed since earlier calculations of Ref.~[\onlinecite{huo_field-induced_2012}] reported a rather different sequence of states in that limit, including the so-called field-induced state. In what follows, we revisit the Bogoliubov-de Gennes (BdG) numerical simulations of Ref.~[\onlinecite{huo_field-induced_2012}] for mesoscopic disks and find that the main claim of their work, i.e. the existence of a field-induced state with winding number $L_+=-1$ in mesoscopic $p$-wave superconductors, is incorrect.  We performed the same type of calculation, under the same conditions, but did not find any signature of the reported behavior. We instead find that the phase diagram with respect to field and temperature is a more conventional one, in agreement with Ginzburg-Landau (GL) simulations of Ref. \onlinecite{huang_phase_2012}.

The taken parameters, same as used in Ref.~[\onlinecite{huo_field-induced_2012}], have been introduced in section~\ref{Ap:subsec:Homo}. First, we show in Fig.~\ref{Ap:Fig.1} the components $\Delta_\pm(r)$ and the supercurrent density $j_\theta(r)$ for states $L_+=-1$ and $L_+=0$. The results for disks with small radii are quite different from those for large radii presented in Fig.~2 of Ref.~[\onlinecite{huo_field-induced_2012}]. As shown in Figs.~\ref{Ap:Fig.1}(a) and (c), the $L_+=-1$ state is time-reversal-symmetric due to boundary effects. As a result, the current is zero in the absence of the magnetic field, which makes the state more stable when $\phi=0$, where $\phi$ is the magnetic flux through the sample. We recover the radial dependence of the supercurrent presented in Ref.~[\onlinecite{huo_field-induced_2012}] only for $R>10\xi_0$, but as shown in the main body of this paper, for those sizes of the sample the cylindrical symmetry of the superconducting state no longer holds. The $L_+=0$ state (shown in Figs.~\ref{Ap:Fig.1}(b) and (d)) has a broken time-reversal symmetry, with spontaneous chiral edge currents. However, the current exhibits sudden changes near the center of the sample due to the competition between the surface current and the cylindrical symmetry.  When $R=1.3\xi_0$, as shown in Fig.~\ref{Ap:Fig.1}(d), this competition, together with the increased importance of quantum confinement induces oscillations in $j_\theta$ on the scale of $1/k_F$, while also enhancing it. $\Delta_+$ is therefore weakened near the center, signaling that the $L_+=0$ state becomes less favorable in small samples.

\begin{figure}[t]
\centering
\includegraphics[width=\columnwidth]{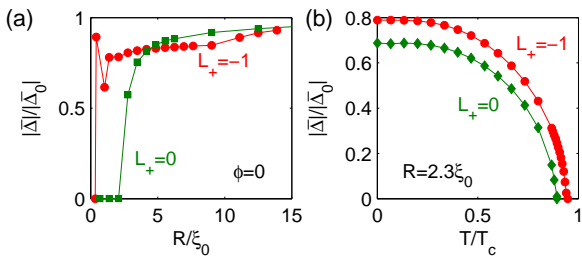}
\caption{(Color online) (a) Spatially averaged order parameter $\bar{\Delta}= \sqrt{|\Delta_+|^2+ |\Delta_-|^2}$ as a function of the sample radius $R$ at $\phi=0$ and $T=0$, for the states $L_+=-1$ and $L_+=0$. (b) $\bar{\Delta}$ as a function of $T$ for sample radius $R=2.3\xi_0$ for $L_+=-1$ at $\phi=0$ and for $L_+=0$ at $\phi=1.5\phi_0$.} \label{Ap:Fig.12}
\end{figure}

In Fig.~\ref{Ap:Fig.12}, we show the calculated dependence of the spatially averaged order parameter $\bar{\Delta}$ for states $L_+=-1$ and $L_+=0$ on the sample size and temperature. This dependence is similar to the results shown in Figs.~3(a) and 3(b) in Ref.~[\onlinecite{huo_field-induced_2012}].  However, the quantum size effect is robust when $R \sim 1/{k_F}$, which results in oscillations in $\bar{\Delta}(R)$. These were not shown in Ref.~[\onlinecite{huo_field-induced_2012}].

\begin{figure}[t]
\centering
\includegraphics[width=\columnwidth]{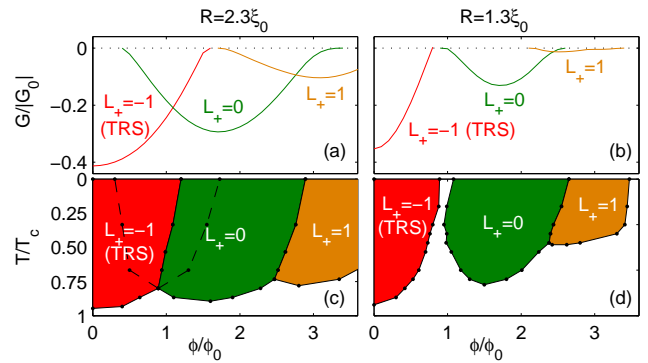}
\caption{(Color online) Free energy density $G$ of the superconducting states as a function of magnetic flux threading the sample, $\phi$, for a $p$-wave mesoscopic disk of radius (a) $R=2.3\xi_0$ and (b) $R=1.3\xi_0$ at $T=0.33T_c$. The corresponding $\phi-T$ equilibrium phase diagrams are shown in (c) and (d), respectively (cf. with the erroneous Figs.~3(c,d) of Ref.~[\onlinecite{huo_field-induced_2012}]).} \label{Ap:Fig.2}
\end{figure}

Once the self-consistent result is reached, the
Gibbs free energy is calculated using Eq.~(\ref{Ap:Eq:Heff_lastep}). The superconducting free energy density $G$, for radii $R=2.3\xi_0$ and $1.3\xi_0$ is shown as a function of $\phi$ in Fig.~\ref{Ap:Fig.2}. These results are in sync with conventional wisdom and similar to those for mesoscopic $s$-wave superconductors obtained by the Ginzburg-Landau (GL) theory in e.g. Ref. \onlinecite{schweigert_vortex_1998}. $G(\phi)$ plots were not presented in Ref.~[\onlinecite{huo_field-induced_2012}], and therefore one has no means to directly validate their $\phi-T$ phase diagram.

From our Fig.~\ref{Ap:Fig.2}(a) and (b) it is clear that at low fields and for the considered dimensions of the sample the ground state is always the $L_+=-1$ state. As the field is increased, a first-order phase transition to the $L_+=0$ state takes place. By decreasing the size of the system, the $L_+=-1$ state becomes more stable while the phase boundary between the two states moves to lower fields. If the size is decreased even further, the $L_+=-1$ and $L_+=0$ superconducting states become separated by the normal state. This finite-size re-entrant effect is similar to already observed oscillations of the critical temperature in $s$-wave superconducting rings \cite{liu_destruction_2001}. This is not surprising since in the small radius limit, both $\Delta_{+}$ and $\Delta_{-}$ components of the order parameter are strongly suppressed in the center of the disk, effectively mimicking a superconducting ring.

Finally, we present in Fig.~\ref{Ap:Fig.2}(c) and (d) the correct ground state $\phi-T$ phase diagrams for samples with radius $R=2.3\xi_0$ and $R=1.3\xi_0$.  The results are physically understandable, as states always appear sequentially in the winding number of the dominant $p_x+ip_y$ component as $\phi$ increases. Moreover, our results at finite temperature are consistent with those shown in Ref.~[\onlinecite{huang_phase_2012}], where a Ginzburg-Landau simulation was performed for a similar system. 

To conclude, we find that the phase diagram of Ref.~[\onlinecite{huo_field-induced_2012}] is incorrect. To substantiate our conclusion we performed identical BdG calculations for a disk geometry and showed explicitly the correct form of the free energy based on which the equilibrium phase diagram is constructed. Our free energy-versus-magnetic field curves provide a clear picture of the nature of the ground state, which at zero-field is always the $L_+=-1$ state (in the range of parameters of interest here, i.e. for small samples). An additional support to our claims are the calculations performed in Ref.~[\onlinecite{huang_phase_2012}] within the GL model, which delivered a very similar phase diagram to ours. Finally we note again that this discussion is related only to the limit of very small samples, whereas the focus of our present paper is on novel states in \textit{larger} mesoscopic $p$-wave samples where cylindrical symmetry is \textit{broken}, leading to a much richer phase diagram.


\section*{Acknowledgments}
This work was supported by the Research Foundation Flanders (FWO-Vlaanderen) and the Special Research Funds of the University of Antwerp.

%


\end{document}